# Derivation of a four-spinor Dirac-like equation for the helium atom and an approximate ground state solution.

**D. L. Nascimento and A. L. A. Fonseca**

Institute of Physics and International Center of Condensed Matter, University of Brasília,

70919-970, Brasília – DF, Brazil,

daniel@fis.unb.br,   alaf@fis.unb.br

**Abstract**

We present a new analysis of the connection between the classical conservation theorems and the role played by the Dirac matrices in order to obtain a four spinor version of the Dirac equation for the two electrons bound problem. The separation of the set of equations in its angular and radial parts is done and an asymptotic solution for the radial system of equations is discussed for the ground state of the atom.

**1. Introduction.**

In a previous work [1] we have shown that, by taken into account the classical conservation theorems  before the Dirac matrices are found, it is possible to use  2x2 matrices instead the usual 4x4 matrices in the solution of the hydrogen atom, which yields a considerable reduction in the complexity of the problem.  In this work, we try to get the same reduction in the dimensions of the Dirac matrices for the problem of the helium atom, which allows us to use 4x4 matrices instead of the 16x16 matrices of the Breit theory of the helium [2,3], which in spite of this is our starting point.

We differ from Breit firstly by not considering the retardation effect in this work and secondly  by trying to use a procedure that preserves the individual identity of each electron. This is done in order to get a truly covariant system of equations, since it is a trivial fact of relativity theory that we can not sum up the geodesics of individual particles. To accomplish this we consider first a single electron plus the nucleus, i.e., the $He^+$ ion, as the background system  and afterwards we introduce the second electron through a penetration parameter which allows the superposition of the individual Hamiltonian equations. This yields a system of differential equations whose  wave function is a solution for the whole system, depending on the separation parameter, which is used at the end of the calculation to obtain the least energy configuration of the system.

In the remaining of the paper we separate the angular and radial parts of the partial differential equations that come from our helium Dirac-like equation and found the angular eigenfunctions that allow to separate the system of radial equations. We then solve this radial equations system asymptotically, from which we get a determination of the ground state energy which agrees to 0.1% with the experimental data and also verify that the exact ion limit is reached when the invader electron goes to infinity [4].

## 1. Theory

In the nucleus rest frame, the relativistic Hamiltonians for the electrons of the helium atom in natural units $\hbar = c = 1$ are

$$H_1 = \sqrt{\mathbf{p}_1^2 + m^2} - \frac{2\alpha}{r_1} + \frac{\alpha}{r_{12}}, \qquad H_2 = \sqrt{\mathbf{p}_2^2 + m^2} - \frac{2\alpha}{r_2} + \frac{\alpha}{r_{12}}, \qquad (1a,b)$$

where $r_{12} = \sqrt{r_1^2 + r_2^2 - 2r_1 r_2 \cos\theta}$.

We now search for a Dirac equation corresponding to the quantization of the classical Eq. (1), in the nucleus rest frame coordinate system in which the motion occurs in the plane defined by the nucleus and the two electrons, i.e., $p_z = 0$, whose z axis may be moving at constant velocity with respect to the z axis of another inertial system, so that the system is invariant against space translations in this direction. The quantization is done in a way similar to that performed by Breit, in which each square root is "linearized" individually:

$$\mathbf{p}_1^2 + m^2 = \left(\gamma^5 p_{1y} - \gamma^3 p_{1x} + m\gamma^0\right)^2, \qquad (2a)$$

$$\mathbf{p}_2^2 + m^2 = \left(\gamma^1 p_{2y} - \gamma^2 p_{2x} + m\gamma^0\right)^2. \qquad (2b)$$

We need of five $4 \times 4$ anticomuting matrices, they are the four usual $\gamma^\mu$ Dirac matrices together with the $\gamma^5$ matrix which always appears connected with Dirac´s theory:

$$\gamma^0 = \begin{pmatrix} 1 & 0 \\ 0 & -1 \end{pmatrix}, \qquad \boldsymbol{\gamma} = \begin{pmatrix} 0 & i\boldsymbol{\sigma} \\ -i\boldsymbol{\sigma} & 0 \end{pmatrix}, \qquad \gamma^5 = -i\gamma^0\gamma^1\gamma^2\gamma^3 = \begin{pmatrix} 0 & 1 \\ 1 & 0 \end{pmatrix}, \qquad (3)$$

where $\sigma_x, \sigma_y, \sigma_z$ are the Pauli spin matrices. As is well known, the $\gamma$ matrices obey $\gamma^\mu \gamma^\nu + \gamma^\nu \gamma^\mu = 2\delta^{\mu\nu}$, for $\mu,\nu = 0,1,2,3,5$, that is, are unitary and anticommute in pairs, as required to make equal the two sides of Eq.(2).

We now define a Hamiltonian function for the whole system which is composed of a inner Hamitonian $H_1$ of the ion $H_e^+$ and an outer Hamiltonian $H_2$, which takes into account the invader electron, which is superposed to the former through a penetration factor $\sigma$, that is, $H = (1-\sigma)H_1 + 2\sigma H_2$. We see that $\sigma = 0$ corresponds to the ion limit when $r_{12} \to \infty$ and the electron 2 is not present; on the other hand, $\sigma = 1$ correspond to the limit when the two electrons form a single system with perfectly symmetric positions so that the system Hamiltonian becomes two times the Hamiltonian of one of the electrons, which was chosen by convenience to be the electron 2. In fact, $\sigma$ is a function of $r_{12}$, so that the equations of the system will be solved for $r_{12}$ constant and then, at the final of the calculation, this constant is varied in order the equilibrium configuration may be obtained.

By using the momenta operators $\mathbf{p}_1 = -i\nabla_1$ and $\mathbf{p}_2 = -i\nabla_2$, the linear Hamiltonian-like matrix operator associated with Eqs.(1) becomes

$$H = (1-\sigma)\left[i\left(\gamma^3 \partial_{x_1} - \gamma^5 \partial_{y_1}\right) - \frac{2\alpha}{r_1}\right] + 2\sigma\left[i\left(\gamma^2 \partial_{x_2} - \gamma^1 \partial_{y_2}\right) - \frac{2\alpha}{r_2}\right] + (1+\sigma)\left(m\gamma^0 + \frac{\alpha}{r_{12}}\right). \quad (4)$$

Now, it may immediately be seen that the total orbital angular momentum operator

$$J_z = iy_1 \partial_{x_1} - ix_1 \partial_{y_1} + iy_2 \partial_{x_2} - ix_2 \partial_{y_2} \quad (5)$$

does not commute with $H$, however $M = J_z + \frac{1}{2}(\alpha_{1z} + \alpha_{2z})$ commutes with H, in which $\alpha_{1,z} = -i\gamma^5 \gamma^3 = \begin{pmatrix} -\sigma_z & 0 \\ 0 & \sigma_z \end{pmatrix}$ and $\alpha_{2,z} = -i\gamma^1 \gamma^2 = \begin{pmatrix} -\sigma_z & 0 \\ 0 & -\sigma_z \end{pmatrix}$ are algebraic complements of $\gamma^\mu$.

The diagonalization problems to be solved are therefore $H\psi = E\psi$ and $M\psi = j\psi$, where $\psi = (\chi_1, \chi_2, \chi_3, \chi_4)$ is a four-spinor. Now, in order to get a truly covariant equation, we left-multiply the energy eigen-problem by $\gamma^0$, so that we get

$$\left[(\phi_{12} - E)\gamma^0 + i(1-\sigma)\left(\gamma^0 \gamma^5 \partial_{x_1} - \gamma^0 \gamma^3 \partial_{y_1}\right) + 2i\sigma\left(\gamma^0 \gamma^1 \partial_{x_2} - \gamma^0 \gamma^2 \partial_{y_2}\right) + (1+\sigma)m\right]\psi = 0, \quad (6)$$

where $\phi_{12} = -\frac{2(1-\sigma)\alpha}{r_1} - \frac{4\sigma\alpha}{r_2} + \frac{(1+\sigma)\alpha}{r_{12}}$ is the total potential energy function. It may immediately be seen that Eq.(6) can be put in the explicit covariant form $\left[(1-\sigma)\zeta_{1\mu}\pi_1^\mu + 2\sigma\zeta_{2\mu}\pi_2^\mu\right]\psi = 0$, by rewriting it in terms of the matrix operators $\zeta_{1\mu} = \left(-1, -\gamma^0\gamma^5, \gamma^0\gamma^3, \gamma^0\right)$, $\zeta_{2\mu} = \left(-1, -\gamma^0\gamma^1, \gamma^0\gamma^2, \gamma^0\right)$ and the effective momentum operators $\pi_k^\mu = \left(m, -i\partial_{x_k}, -i\partial_{y_k}, -\frac{2\alpha}{r_k} + \frac{\alpha}{r_{12}} - E\right)$ for $k = 1, 2$.

Explicitly, Eq.(6) becomes the following system of linear partial differential equations

$$q_+\chi_1 - (1-\sigma)\left(\partial_{x_1} + i\partial_{y_1}\right)\chi_3 + 2\sigma\left(-\partial_{x_2} + i\partial_{y_2}\right)\chi_4 = 0$$

$$q_+\chi_2 + (1-\sigma)\left(\partial_{x_1} - i\partial_{y_1}\right)\chi_4 - 2\sigma\left(\partial_{x_2} + i\partial_{y_2}\right)\chi_3 = 0$$

$$q_-\chi_3 + (1-\sigma)\left(-\partial_{x_1} + i\partial_{y_1}\right)\chi_1 + 2\sigma\left(-\partial_{x_2} + i\partial_{y_2}\right)\chi_2 = 0, \qquad (7)$$

$$q_-\chi_4 + (1-\sigma)\left(\partial_{x_1} + i\partial_{y_1}\right)\chi_2 - 2\sigma\left(\partial_{x_2} + i\partial_{y_2}\right)\chi_1 = 0$$

in which the new potential functions $q_\pm = (1+\sigma)m \pm (\phi_{12} - E)$ are introduced for shortness.

Now, in order to separate the angular part of Eq.(7), we must address the angular momentum problem, $\left(\partial_{\theta_1} + \partial_{\theta_2}\right)\psi = i(j \pm 1)\psi$, in which $j = j_1 + j_2$, written in the polar coordinates of the electrons 1 and 2, where $j \pm 1$ are the possible variations of the angular momentum coming from the matrices in $J_z$. The most general forms for the solutions of the angular equation are the set of eigenfunctions

$$\chi_k = f_k e^{i\Phi_k}, \quad f_k = f_k\left(r_1, r_2, \rho(r_1, r_2, \theta)\right), \quad k = 1,...,4, \qquad (8)$$

where the phase functions are $\Phi_1 = \left(j_1 + \tfrac{1}{2}\right)\theta_1 - \left(j_2 + \tfrac{1}{2}\right)\theta_2$, $\Phi_2 = \left(j_1 - \tfrac{1}{2}\right)\theta_1 + \left(j_2 + \tfrac{1}{2}\right)\theta_2$, $\Phi_3 = \left(j_1 - \tfrac{1}{2}\right)\theta_1 + \left(j_2 - \tfrac{1}{2}\right)\theta_2$, $\Phi_4 = \left(j_1 + \tfrac{1}{2}\right)\theta_1 - \left(j_2 - \tfrac{1}{2}\right)\theta_2$, in which $\rho = r_{12}$ and $\theta = \theta_1 - \theta_2$. Formally, the substitution of $\chi_k$ in Eq.(7) makes all complex phases and angular variables vanish and yields a new set of linear equations depending only on the radial variables $r_1, r_2, \rho$; however, in this work we shall limit ourselves to search for solutions satisfying the constraint $\rho(r_1, r_2, \theta) = \rho_0 =$ constant, so that $f_k = f_k(r_1, r_2)$ which simplify considerably the resulting equations and also yields the variational relation $\sigma = g(\rho_0)$ that will be used to get the equilibrium configuration of the system. The role of $\rho$ as a truly variable will be considered in a numerical version of this work which is in development at the moment. Thus we get

$$q_+f_1 - (1-\sigma)\left(\frac{\partial f_3}{\partial r_1} - \frac{j_1 - \tfrac{1}{2}}{r_1}f_3\right) - 2\sigma\left(\frac{\partial f_4}{\partial r_2} - \frac{j_2 - \tfrac{1}{2}}{r_2}f_4\right) = 0$$

$$q_+f_2 + (1-\sigma)\left(\frac{\partial f_4}{\partial r_1} + \frac{j_1 + \tfrac{1}{2}}{r_1}f_4\right) - 2\sigma\left(\frac{\partial f_3}{\partial r_2} + \frac{j_2 - \tfrac{1}{2}}{r_2}f_3\right) = 0$$

$$q_-f_3 - (1-\sigma)\left(\frac{\partial f_1}{\partial r_1} + \frac{j_1 + \tfrac{1}{2}}{r_1}f_1\right) - 2\sigma\left(\frac{\partial f_2}{\partial r_2} + \frac{j_2 + \tfrac{1}{2}}{r_2}f_2\right) = 0, \qquad (9)$$

$$q_-f_4 + (1-\sigma)\left(\frac{\partial f_2}{\partial r_1} - \frac{j_1 - \tfrac{1}{2}}{r_1}f_2\right) - 2\sigma\left(\frac{\partial f_1}{\partial r_2} + \frac{j_2 + \tfrac{1}{2}}{r_2}f_1\right) = 0.$$

In this work we shall restrict ourselves to get the simplest solution for Eq.(9), that is

$$f_k = a_{k00} r_1^{s_1} r_2^{s_2} e^{-\beta_1 r_1 - \beta_2 r_2}, \qquad k = 1, 2, 3, 4, \qquad (10)$$

because the approximation made above restrict heavily the possibility of obtaining energy substates, which depend strongly on higher degree power series coefficients. The substitution of Eq.(10) into Eq.(9) and canceling of the exponential terms yields the new set of equations

$$\left[\gamma_{2\rho} - \frac{2\alpha(1-\sigma)}{r_1} - \frac{4\alpha\sigma}{r_2}\right]a_{100} + (1-\sigma)\left(\beta_1 + \frac{j_1 - s_1 - \frac{1}{2}}{r_1}\right)a_{300} + 2\sigma\left[\beta_2 + \frac{j_2 - s_2 - \frac{1}{2}}{r_2}\right]a_{400} = 0$$

$$\left[\gamma_{2\rho} - \frac{2\alpha(1-\sigma)}{r_1} - \frac{4\alpha\sigma}{r_2}\right]a_{200} + 2\sigma\left[\beta_2 - \frac{j_2 + s_2 + \frac{1}{2}}{r_2}\right]a_{300} + (1-\sigma)\left(\frac{j_1 + s_1 + \frac{1}{2}}{r_1} - \beta_1\right)a_{400} = 0 \qquad (11)$$

$$\left[\gamma_{1\rho} + \frac{2\alpha(1-\sigma)}{r_1} + \frac{4\alpha\sigma}{r_2}\right]a_{300} + 2\sigma\left[\beta_2 + \frac{j_2 - s_2 - \frac{1}{2}}{r_2}\right]a_{200} + (1-\sigma)\left(\beta_1 - \frac{j_1 + s_1 + \frac{1}{2}}{r_1}\right)a_{100} = 0$$

$$\left[\gamma_{1\rho} + \frac{2\alpha(1-\sigma)}{r_1} + \frac{4\alpha\sigma}{r_2}\right]a_{400} - (1-\sigma)\left(\beta_1 + \frac{j_1 - s_1 - \frac{1}{2}}{r_1}\right)a_{200} + 2\sigma\left[\beta_2 - \frac{j_2 + s_2 + \frac{1}{2}}{r_2}\right]a_{100} = 0,$$

where $\gamma_{1\rho} = (1+\sigma)m + E - \dfrac{(1+\sigma)\alpha}{\rho}$ and $\gamma_{2\rho} = (1+\sigma)m - E + \dfrac{(1+\sigma)\alpha}{\rho}$. Now we start with the determination of the coefficients and parameters by observing that the system of equations formed by each negative power $1/r_1$ and $1/r_2$, must vanish separately, in order the coefficients $a_{k00}$ do not vanish:

$$-2\alpha a_{100} + (j_1 - s_1 - \tfrac{1}{2})a_{300} = 0$$
$$-2\alpha a_{200} + (j_1 + s_1 + \tfrac{1}{2})a_{400} = 0$$
$$2\alpha a_{300} - (j_1 + s_1 + \tfrac{1}{2})a_{100} = 0, \qquad (12)$$
$$2\alpha a_{400} - (j_1 - s_1 - \tfrac{1}{2})a_{200} = 0$$

$$-4\alpha a_{100} + (j_2 - s_2 - \tfrac{1}{2})a_{400} = 0$$
$$-4\alpha a_{200} - (j_2 + s_2 + \tfrac{1}{2})a_{300} = 0$$
$$4\alpha a_{300} + (j_2 - s_2 - \tfrac{1}{2})a_{200} = 0, \qquad (13)$$
$$4\alpha a_{400} - (j_2 + s_2 + \tfrac{1}{2})a_{100} = 0.$$

For this condition be fulfilled, it is necessary that the determinant of the systems Eqs(12-13) vanish, from which we get the relations $s_1 = -\frac{1}{2} + \sqrt{j_1^2 - 4\alpha^2}$ and $s_2 = -\frac{1}{2} + \sqrt{j_2^2 - 4\alpha^2}$.

Now we get back to the original system Eq.(11), let the indices of the coefficients vary by one unity, assume that $a_{k\mu\nu} = a_{k\nu\mu}$, for $\mu \neq \nu = 0,1$, since these coefficients are in fact null, and equate the system of equations for the same powers, from which we get the virtual recurrence system of equations $R_1 = 0, R_2 = 0, R_3 = 0, R_4 = 0$, in which the recurrence functions are

$$R_1 = \gamma_{2\rho} a_{100} - 2\alpha(1+\sigma) a_{110} - (1-\sigma)\left(1 + \sqrt{j_1^2 - 4\alpha^2} - j_1\right) a_{310}$$
$$+ (1-\sigma)\beta_1 a_{300} - 2\sigma\left(1 + \sqrt{j_2^2 - 4\alpha^2} - j_2\right) a_{410} + 2\sigma\beta_2 a_{400}$$
$$R_2 = \gamma_{2\rho} a_{200} - 2\alpha(1+\sigma) a_{210} - 2\sigma\left(1 + \sqrt{j_2^2 - 4\alpha^2} + j_2\right) a_{310} + 2\sigma\beta_2 a_{300} + (1-\sigma)\left(1 + \sqrt{j_1^2 - 4\alpha^2} + j_1\right) a_{410} + (1-\sigma)\beta_1 a_{400}$$

$$\quad (14)$$

$$R_3 = \gamma_{1\rho} a_{300} + 2\alpha(1+\sigma) a_{310} - (1-\sigma)\left(1 + \sqrt{j_1^2 - 4\alpha^2} + j_1\right) a_{110} + (1-\sigma)\beta_1 a_{100} - 2\sigma\left(1 + \sqrt{j_2^2 - 4\alpha^2} - j_2\right) a_{210} + 2\sigma\beta_2 a_{200}$$

$$R_4 = \gamma_{1\rho} a_{400} + 2\alpha(1+\sigma) a_{410} - 2\sigma\left(1 + \sqrt{j_2^2 - 4\alpha^2} + j_2\right) a_{110} + 2\sigma\beta_2 a_{100} + (1-\sigma)\left(1 + \sqrt{j_1^2 - 4\alpha^2} - j_1\right) a_{210} - (1-\sigma)\beta_1 a_{200}.$$

We can immediately see that, since the coefficients $a_{k\mu\nu}$ must vanish for $\mu \neq \nu = 0,1$, then the determinant of the coefficients $a_{k00}$ must vanish also, so that the remaining system have a non trivial solution, that is

$$\begin{vmatrix} \gamma_{2\rho} & 0 & (1-\sigma)\beta_1 & 2\sigma\beta_2 \\ 0 & \gamma_{2\rho} & 2\sigma\beta_2 & -(1-\sigma)\beta_1 \\ (1-\sigma)\beta_1 & 2\sigma\beta_2 & \gamma_{1\rho} & 0 \\ 2\sigma\beta_2 & -(1-\sigma)\beta_1 & 0 & \gamma_{1\rho} \end{vmatrix} = 0, \quad (15)$$

This determinant is in fact an eigenvalue equation for $\beta_1$ as a function of $\beta_2$ and the other parameters. The non negative solution we search for is $\beta_1 = \dfrac{\sqrt{\gamma_{1\rho}\gamma_{2\rho} - 4\sigma^2 \beta_2^2}}{1-\sigma}$. Now, the homogeneous system of equations in the coefficients $a_{k00}$ corresponding to the determinant Eq.(15) has a kernel generated by the eigenvalue $\beta_1$, whose base is given by the two linearly independent column vectors

$$\psi_1 = \begin{pmatrix} -(1-\sigma)\beta_1/\gamma_{2\rho} \\ -2\sigma\beta_2/\gamma_{2\rho} \\ 1 \\ 0 \end{pmatrix}, \quad \psi_2 = \begin{pmatrix} -2\sigma\beta_2/\gamma_{2\rho} \\ -(1-\sigma)\beta_1/\gamma_{2\rho} \\ 0 \\ 1 \end{pmatrix}, \quad (16)$$

out of which $\psi = a_{300}\psi_1 + a_{400}\psi_2$ is a general kernel vector. These vector basis generates by its turn relations among the power coefficients given by $a_{100} = -\frac{(1-\sigma)\beta_1}{\gamma_{2\rho}}a_{300}$, $a_{200} = -\frac{2\sigma\beta_2}{\gamma_{2\rho}}a_{300}$ for the former vector and $a_{100} = -\frac{2\sigma\beta_2}{\gamma_{2\rho}}a_{400}$, $a_{200} = -\frac{(1-\sigma)\beta_1}{\gamma_{2\rho}}a_{400}$ for the later. The last step in order to be able to make the evaluation of the energy eigenvalue of the system is obtained by first forming the null line vector $R = [R_1, R_2, R_3, R_4]$ and then making a contraction of it with one of the kernel vectors. Since it may be seen that both kernel vectors produce the same energy eigenvalue, so that the solutions in $a_{300}$ and $a_{400}$ are degenerated, we chose to make the contraction with the first kernel vector, that is, $R\psi_1 = 0$. This operation, as expected, eliminates the coefficients $a_{k00}$, but it maintains the coefficients $a_{k10}$, that is

$$\left[\frac{2\alpha(1-\sigma^2)\beta_1}{\gamma_{2\rho}} - (1-\sigma)\left(1+\sqrt{j_1^2 - 4\alpha^2} + j_1\right)\right]a_{110} + 2\sigma\left[\frac{2\alpha(1+\sigma)\beta_2}{\gamma_{2\rho}} - 1 - \sqrt{j_2^2 - 4\alpha^2} + j_2\right]a_{210} + \\ +\left[\left(1+\sqrt{j_1^2 - 4\alpha^2} - j_1\right)\frac{(1-\sigma)^2\beta_1}{\gamma_{2\rho}} + 4\sigma^2\beta_2\left(1+\sqrt{j_2^2 - 4\alpha^2} + j_2\right) + 2\alpha(1+\sigma)\right]a_{310} = 0 \quad (17)$$

Now we decrease the indices μ by one step, in order we can obtain another relation for the coefficients $a_{k00}$, and we use the first contraction to eliminate the coefficients $a_{k00}$, that is,

$$-\left[\frac{2\alpha(1-\sigma^2)\beta_1}{\gamma_{2\rho}} - (1-\sigma)\left(1+\sqrt{j_1^2 - 4\alpha^2} + j_1\right)\right](1-\sigma)\frac{\beta_1}{\gamma_{2\rho}} - 4\sigma^2\left[\frac{2\alpha(1+\sigma)\beta_2}{\gamma_{2\rho}} - 1 - \sqrt{j_2^2 - 4\alpha^2} + j_2\right]\frac{\beta_2}{\gamma_{2\rho}} + \\ +\left(1+\sqrt{j_1^2 - 4\alpha^2} - j_1\right)\frac{(1-\sigma)^2\beta_1}{\gamma_{2\rho}} + 4\sigma^2\beta_2\left(1+\sqrt{j_2^2 - 4\alpha^2} + j_2\right) + 2\alpha(1+\sigma) = 0. \quad (18)$$

Now, we write by convenience $\beta_2 = h\beta_1$, where $h$ is to be determined below, then, after a little of algebra, we get the fundamental relation $\beta_1 = \frac{\alpha(1+\sigma)(\gamma_{1\rho} - \gamma_{2\rho})}{(1-\sigma^2)\sqrt{j_1^2 - 4\alpha^2} + 4\sigma^2 h\left(1+\sqrt{j_2^2 - 4\alpha^2}\right)}$ among the electron parameters. This relation allow us, by substituting the expressions for γ1ρ, γ2ρ and $\beta_1$ found above, to get a first expression for the energy eigenvalues we are searching for

$$E = \frac{\alpha(1+\sigma)}{\rho_0} + \frac{(1+\sigma)m}{\sqrt{1 + \frac{4\alpha^2(1+\sigma)^2\left[(1-\sigma)^2 + 4\sigma^2 h^2\right]}{(1-\sigma^2)\sqrt{j_1^2 - 4\alpha^2} + 4\sigma^2 h\left(1+\sqrt{j_2^2 - 4\alpha^2}\right)}}}. \quad (19)$$

However, this is not yet the final point, since we need still to found out a value for $h$ and the connection between σ and $\rho_0$, so that we can obtain a numerical evaluation of the atom energy. This is done by recalling a relation we have stated in a paper where we discussed the connection between orbital properties and the parameters of the wave functions. From there come the relations $r_{10} = s_1/\beta_1$, $r_{20} = s_2/\beta_2$ among the stable orbital radii and the points of maxima of the radial part of the wave equation.

We have further assumed that at the equilibrium configuration we have $\rho_0 = r_{10} + r_{20}$ and also the linear connection $r_{10} = \sigma r_{20}$ between the electron equilibrium radii. From these relations we finally get $h = \sigma s_2/s_1$, and together with Eq.(19), also a determination for the equilibrium distance between the electrons $\rho_0 = \dfrac{C_1}{2\sigma m\alpha(1+\sigma)}$. (We do not try to obtain the inverse function $\sigma = \sigma(\rho_0)$, because this process would not bring anything new, but a complicated algebraic function). We have defined for shortness, besides the parameter C1($\sigma$), also another parameter C2($\sigma$) to be used below as

$$C_1 = \sqrt{\left[(1-\sigma)^2(s_1+\tfrac{1}{2})s_1 + 4\sigma^3(s_2+\tfrac{3}{2})s_2\right]^2 + 4\alpha^2(1+\sigma)^2\left[(1-\sigma)^2 s_1^2 + 4\sigma^4 s_2^2\right]}$$

$$C_2 = \sqrt{1 + \dfrac{4\alpha^2(1+\sigma)^2\left[(1-\sigma)^2 s_1^2 + 4\sigma^4 s_2^2\right]}{\left[(1-\sigma)^2(s_1+\tfrac{1}{2})s_1 + 4\sigma^3(s_2+\tfrac{3}{2})s_2\right]^2}} \,. \tag{20}$$

At this point we have finally fulfilled all the steps toward to get an expression for the energy eigenvalues in terms of the basic electron properties besides the variation factor $\sigma$, that is

$$E = \dfrac{2\sigma m\alpha^2(1+\sigma)^2}{C_1} + \dfrac{(1+\sigma)m}{C_2} \,, \tag{21}$$

and the equilibrium radii becomes then $r_{10} = \dfrac{\sigma}{1+\sigma}\rho_0$ and $r_{20} = \dfrac{\rho_0}{1+\sigma}$.

At this point we can make a plot of the energy excess $\Delta E = E - (1+\sigma)m$ of the system against the effective mass $(1+\sigma)m$. After considering the unit conversion factors Hartree = $m\alpha^2$ and the Bohr radius $a_0 = 1/(m\alpha)$, we get the non-dimensional energy and radii equations

$$\Delta E_{an} = \dfrac{2\sigma(1+\sigma)^2}{C_1} + \left[\dfrac{(1+\sigma)}{C_2} - 1 - \sigma\right]\dfrac{1}{\alpha^2} \,, \tag{22}$$

$$\rho_0 = \dfrac{C_1}{2\sigma(1+\sigma)} \,. \tag{23}$$

The plot of the energy as a function of $\sigma$ is shown in fig1, from which we immediately see that the ion ground state limit $\Delta E = -2$ occurs for $\sigma = 0$. The minimum of the energy for $j_1 = j_2 = 1$ corresponds to the inner orbital state for the parahelium atom (or the state 1s-1s of the Spectroscopy). From the plot we see that the equilibrium value occurs approximately for $\sigma_0 = 0.1775$, for which value we get an energy ground state of $\Delta E = -2.90589$, which less than 0.1% below the experimental value of $\Delta E_{Exp} = -2.90330$, which means that our determination of the ground state energy of the atom, although the rough approximations done in order we could get this first approximation solution. The equilibrium radii found $r_{10} = 0.130$, $r_{20} = 0.732$ and $\rho_0 = 0.862$ seems also in agreement with the corresponding experimental determinations.

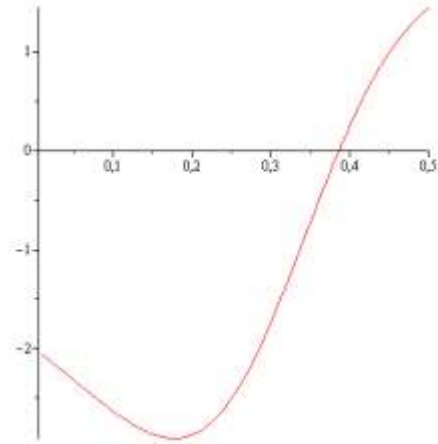

Fig1. Excess energy variation against σ.

**Conclusion.**

In this work we have derived a four-spinor Dirac-like equation for the helium. We have discussed the general structure of the equations, found the matrices needed, separated the system of equations, found the angular eigenfunctions that decouple the system and solutions for the radial equation, in the approximation of constant inter-electron distance. This has made possible to calculate the ground state energy eigenvalue which agrees with the experimental data. We have also shown that the ion helium atom ground state is obtained as a limit case, as expected.

**References.**

[1] D.L. Nascimento and A.L.A.Fonseca, *A 2D spinless version of Dirac's equation written in a noninertial frame of reference,* Int. J. Q. Chem. **111** 1361-69 (2011).

[2] G. Breit, *The effect of retardation on the interaction of two electrons*, Phys. Rev. **34**, 553-573 (1929).

[3] H.A. Bethe, E.E. Salpeter, *Quantum Mechanics of One- and Two-Electron Atoms*, Plenum Press, New York 1977.

[4] D.L. Nascimento, A.L.A.Fonseca, F. F. Monteiro and M. A. Amato, *A Variational Approach for Numerically Solving the Two-Component Radial Dirac Equation for One-Particle Systems,* J. Mod. Phys. **3**, 350-354 (2012).